\documentclass[conference]{IEEEtran}
\IEEEoverridecommandlockouts

\usepackage{amsmath,amsfonts}
\usepackage{url}
\usepackage{cite}
\usepackage{stmaryrd}
\usepackage{graphicx}
\DeclareGraphicsExtensions{.pdf,.jpeg,.png}

\begin{document}

\title{Leveraging Code Automorphisms for Improved Syndrome-Based Neural Decoding\\
\thanks{This work has been funded by the French National Research Agency AI4CODE project (Grant ANR-21-CE25-0006). Part of it was performed using HPC resources from GENCI-IDRIS (Grant 2025-AD011016057)}
}
\author{\IEEEauthorblockN{Raphaël Le Bidan, Ahmad Ismail, Elsa Dupraz, Charbel Abdel Nour}
\IEEEauthorblockA{\textit{IMT Atlantique, Lab-STICC UMR CNRS 6285, Brest, France} \\
}
}

\maketitle

\begin{abstract}
Syndrome-based neural decoding (SBND) has emerged as a promising deep learning approach for soft-decision decoding of high-rate, short-length codes. However, this approach still has substantial room for improvement. In this paper, we show how to leverage code automorphisms to enhance the ability of existing SBND models to learn and generalize through data augmentation during training and inference. As a result, for the short high-rate codes considered, we obtain models that closely approach MLD performance using small datasets and proper training. Our findings also suggest that many prior results for SBND models in the literature underestimate their true correction capability due to undertraining.
\end{abstract}


\section{Introduction}
\IEEEPARstart{T}{his} work addresses the problem of soft-decision decoding for generic  linear block codes of short length. It is well established that the optimal decoder minimizing the decoding error probability is the maximum likelihood decoder (MLD), which is computationally prohibitive for most codes, even short ones. The challenge thus lies in closely approximating its performance with the simplest possible decoder. This long-standing problem in coding theory has gained renewed interest with ultra-reliable low-latency communications (URLLC) requirements in B5G and 6G wireless networks \cite{yue2023efficient}.

Ordered Statistics Decoding (OSD) \cite{fossorier1995soft} is one possible solution. Belief Propagation (BP) decoding \cite{shen2025toward}, successive cancellation list decoding \cite{lin2025toward}, and guessing decoding \cite{wang2025ordered} are compelling alternatives. However, each of these solutions comes with its own trade-offs. To date, no universal decoding approach has been found that achieves a satisfactory balance between error correction performance and hardware complexity. In this context, deep learning-based decoding is a promising direction worth exploring, with potential for performance gains and possibly new algorithmic insights.

Most contributions on neural decoding of error-correcting codes fall into two categories: model-based and model-free approaches. Model-based decoders enhance existing algorithms by augmenting them with learned prediction layers or functions to address their weaknesses. A prominent example is neural BP decoding \cite{nachmani2016learning}, which introduces learnable weights in message-passing computations to mitigate the detrimental effects of short cycles. While neural augmentation does improve performance, the gains often remain modest. This paper focuses on the second family: model-free neural decoders that use deep neural networks (DNN) to approximate MLD through a data-driven approach with minimal inductive bias \cite{gruber2017deep}. 

The impetus for this work was to assess how closely model-free decoders can approach MLD performance for short high-rate codes. This requires identifying suitable DNN architectures and training them to their full capability. We therefore focus on syndrome-based neural decoding (SBND) \cite{bennatan2018deep} as the most promising approach within the model-free decoder family. However, this approach still has substantial room for improvement. In particular, existing SBND models exhibit computational costs and memory requirements that significantly exceed those of classical decoders. Another concern is their reported frame error rate (FER) performance, which shows a substantial gap to MLD on many codes \cite{yuan2025design}, despite the fact that prior work on SBND has largely focused on improving performance through novel architectures (see, e.g., \cite{choukroun2022error,debonirovella2023improved,choukroun2024foundation,lau2025interplay,cohen2025hybrid}). As pointed out in \cite{ismail2025doing}, the limited performance of these models can be partly attributed to their training methodology. Ref. \cite{ismail2025doing} also demonstrated that superior performance can be achieved by training on carefully designed fixed datasets rather than on-demand data. 

The present paper continues this effort to train existing SBND models to their full potential. Our contribution is twofold. First, we propose to introduce data augmentation into the training and show how to leverage code automorphisms for this purpose. This allows us to extract the maximum amount of information from datasets and train SBND models that learn and generalize better. Second, we show how code automorphisms can be reused at inference through test-time augmentation (TTA), rediscovering automorphism ensemble decoding \cite{geiselhart2021automorphism} through the lens of deep learning. Overall, by combining principles from coding theory with best practices from deep learning, we demonstrate on standard benchmark codes from the SBND literature that existing models can closely approach MLD performance using very small datasets with proper training. 

The paper is organized as follows. Section \ref{seq:sbnd} introduces the syndrome-based neural decoding framework, reviews the main models from the literature, and outlines the training process. Section \ref{seq:train_time_aug} describes how to use code automorphisms for data augmentation during training. Section \ref{seq:test_time_aug} extends the same principle to improve model performance at inference time. The practical benefits of the proposed data augmentation methods are demonstrated through a series of experiments presented in Section \ref{seq:experiments}. Section \ref{seq:conclusion} concludes this work. 

\section{Syndrome-based neural decoding}
\label{seq:sbnd}

\subsection{Maximum-likelihood soft-decision decoding}

Consider a binary linear block code $\mathcal{C}$ of length $n$ and dimension $k$, with parity-check matrix $\mathbf{H}$. We denote by $\mathbb{F}_2^n$ the $n$-dimensional vector space over the binary Galois field $\mathbb{F}_2=\{0,1\}$. Let $\mathbf{y}=(y_1,\ldots,y_n) \in \mathbb{R}^n$ be the noisy observation corresponding to the transmission of a codeword $\mathbf{c}=(c_1,\ldots,c_n)$ from $\mathcal{C}$ over an additive white Gaussian noise (AWGN) channel using binary phase-shift keying (BPSK) modulation. Soft-decision decoding uses $\mathbf{y}$ to infer the transmitted codeword. It is well established that MLD minimizes the codeword error probability and that the soft-decision MLD rule can be implemented in various ways. The formulation of interest here involves making a symbol-by-symbol hard decision $\mathbf{z}=(z_1,\ldots,z_n) \in \mathbb{F}_2^n$ on $\mathbf{y}$ by thresholding its components, computing the syndrome $\mathbf{s}=\mathbf{z}\mathbf{H}^t \in \mathbb{F}_2^{n-k}$, and then searching for the error pattern $\mathbf{e}_\text{ML} \in \mathbb{F}_2^n$ that minimizes the cost function $w(\mathbf{e}) = \sum_{i:e_i=1} |y_i|$ among the $2^k$ candidate error patterns in the coset associated with syndrome $\mathbf{s}$ \cite{snyders1989maximum}. It follows that the input pair $(|\mathbf{y}|, \mathbf{s})$ constitutes a sufficient statistic for MLD, where $|\mathbf{y}|$ is a shorthand notation for the bit-reliability vector $(|y_1|,\ldots,|y_n|)$. However, for most linear codes, finding the most likely error pattern $\mathbf{e}_\text{ML}$ is computationally intractable. Unlike hard-decision decoding, it cannot be precomputed and tabulated for each syndrome value since the cost function $w(\mathbf{e})$ to minimize depends on $\mathbf{y}$.

\begin{figure}[!t]
\centering
\includegraphics[width=88mm]{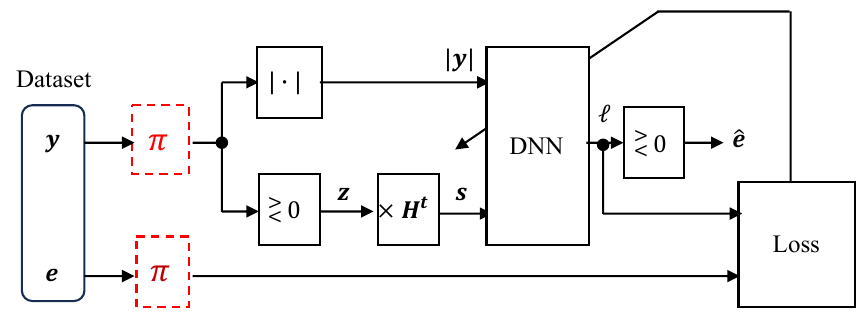}
\caption{Syndrome-based neural decoder architecture and supervised training setup. Data augmentation is introduced through the optional red dotted boxes.}
\label{fig:sbnd}
\end{figure}

\subsection{Principle of SBND}

First introduced in \cite{bennatan2018deep}, SBND uses a DNN to approximate MLD. As illustrated in Fig. \ref{fig:sbnd}, the DNN predicts the most likely error pattern $\mathbf{e}_\text{ML}$ directly from the input pair $(|\mathbf{y}|, \mathbf{s})$, thereby avoiding the exhaustive search required by MLD.  Note that the syndrome value is part of the model input. Since the syndrome space grows exponentially with the number of parity bits, \emph{SBND is best suited to high-rate codes of short-to-moderate length with small redundancy $n-k$}, as is the MLD formulation from which it derives. In practice, the model output is a logit vector $\boldsymbol{\ell} \in \mathbb{R}^n$. The sign of $\ell_i$ provides the hard decision $\hat{e}_i$ on each bit of the predicted error pattern $\hat{\mathbf{e}}$. 

\subsection{DNN models for error pattern prediction}

Early work on SBND showed stacked Gated Recurrent Units (GRU) to be superior to multilayer perceptrons \cite{bennatan2018deep}. Subsequently, the error-correction code transformer (ECCT) was introduced in \cite{choukroun2022error} as an alternative sequence-based SBND model built on the Transformer architecture. Various enhancements to ECCT have been proposed to improve performance \cite{park2025crossmpt,lau2025interplay,cohen2025hybrid}, reduce complexity \cite{park2025crossmpt}, and broaden applicability \cite{choukroun2024foundation}. Hereafter, we only present results for the original ECCT, as it has become a common baseline in the literature.

The ECCT consists of an encoder-only Transformer augmented with a linear classification head. The input pair $(|\mathbf{y}|, \mathbf{s})$ is first converted into a sequence of $2n-k$ learnable embedding vectors in $\mathbb{R}^d$, where $d$ is the embedding dimension. This input representation passes through $L$ successive Transformer blocks, each comprising a masked bidirectional self-attention layer with $h$ attention heads, followed by a feedforward network. This produces a latent representation in $\mathbb{R}^{(2n-k) \times d}$ of the error pattern, which is projected in two steps onto the output logit vector $\boldsymbol{\ell} \in \mathbb{R}^n$ by the linear classification head. We refer the reader to \cite{choukroun2022error} for details. No recipe yet exists for choosing parameters $d$, $L$, and $h$, the first two being the most important for model performance (in that order). 

\subsection{Training procedure}

SBND models are trained in a supervised manner. Most prior work uses the true channel error patterns $\mathbf{e}=\mathbf{z}-\mathbf{c}$ as labels, with on-demand data generation, wherein new training batches are generated at random at each step (see, e.g., \cite{bennatan2018deep,debonirovella2023improved,choukroun2024foundation,lau2025interplay,cohen2025hybrid}). Hereafter, unless otherwise specified, we adopt the approach advocated in \cite{ismail2025doing} and use fixed datasets for training with maximum-likelihood (ML) error patterns as labels. In the following, we refer to these two types of supervision targets as \emph{oracle labels} and \emph{ML labels}, respectively. The training setup is illustrated in Fig.~\ref{fig:sbnd}. A dataset consists of a finite number of training samples $\{(|\mathbf{y}^{(i)}|, \mathbf{s}^{(i)}); \mathbf{e}_{\text{ML}}^{(i)}\}_{i=1}^M$. It is constructed by BPSK transmission of the all-zero codeword to obtain the noisy observations $\mathbf{y}$, followed by MLD decoding to get the corresponding ML error patterns $\mathbf{e}_{\text{ML}} = \mathbf{z}-\mathbf{c}_{\text{ML}}$.
MLD is implemented using an OSD decoder of sufficient order $i_\text{max}=\lceil d_\text{min}/4 \rceil$, where $d_\text{min}$ denotes the minimum distance of $\mathcal{C}$ \cite{fossorier1995soft}. Regardless of whether training relies on a fixed dataset or on-demand data generation, the model is always trained exclusively on observations $\mathbf{y}$ with non-zero syndrome $\mathbf{s}$, since decoding would be skipped otherwise in practice.

Soft-decision decoding of an error-correcting code is a multi-class classification problem. In principle, an SBND model should be trained to minimize the cross-entropy loss, but the prohibitive number of classes, $2^k$, precludes this approach. The usual solution is to decompose the decoding of the received vector into $n$ binary classification problems and minimize the resulting average binary cross-entropy:
\begin{equation*}
\mathcal{L}(\boldsymbol{\ell}, \mathbf{e}) = - \frac{1}{n} \sum_{i=1}^n [ e_i \log(\sigma(\ell_i)) + (1-e_i) \log (1-\sigma(\ell_i))] 
\end{equation*}
where $\sigma(x)$ denotes the sigmoid function. But doing so ignores the bit dependencies induced by the code. As a result, the decoder tends to deviate slightly from the objective of correctly predicting the entire error pattern, in favor of better classifying each bit individually. SBND models trained this way typically achieve very good bit error rates (BER) after decoding while their FER performance lags further behind, especially compared to MLD.

\section{Leveraging Code Automorphisms for Data Augmentation at Training}
\label{seq:train_time_aug}

Constructing the training set can be computationally expensive when dealing with powerful codes having large minimum distance $d_\text{min}$, even with reduced-complexity decoders such as OSD. We therefore want to train the SBND model on a small number of examples to minimize the cost of dataset construction, and possibly also the storage requirements, while maximizing the value extracted from available samples to avoid penalizing model performance. When training a model from limited data, data augmentation is a common practice in deep learning. It consists in applying a randomly selected transformation from a predefined set to each training example. This exposes the model to greater example diversity, strengthening its ability to generalize to unseen samples. The choice of appropriate transformations depends on both the nature of the data and the learning task. In the context of SBND, we propose to use code automorphisms for this purpose.

\subsection{Code automorphism group}

Let $\pi$ be a permutation of the set of integers $\llbracket 1,n\rrbracket$ and $\pi(\mathbf{c})$ the vector resulting from applying this permutation to the coordinates of vector $\mathbf{c}$. A coordinate permutation mapping each codeword of $\mathcal{C}$ to another codeword, not necessarily distinct from the original, defines an automorphism of $\mathcal{C}$. The set of code automorphisms forms a group denoted 
$\text{Aut}(\mathcal{C}) = \left\{ \pi: \pi(\mathbf{c}) \in \mathcal{C}\; \forall \mathbf{c} \in \mathcal{C} \right\}.$ The complete structure of $\text{Aut}(\mathcal{C})$ is known only for a small number of codes. 

The example codes in this paper are primitive binary Bose--Chaudhuri--Hocquenghem (BCH) codes and Polar codes. The automorphism group of a primitive binary BCH code of length $n=2^m-1$ includes the subgroup of cyclic shifts $\{\pi_s: i \mapsto (i+s)\,\text{mod}\,n, i \in \llbracket 1,n\rrbracket \}_{s \in \llbracket 0,n-1\rrbracket}$ and the subgroup of Frobenius permutations $\{\pi_j: i \mapsto i 2^j\,\text{mod}\,n, i \in \llbracket 1,n\rrbracket \}_{j \in \llbracket 0,m-1\rrbracket}$, yielding a minimum of $m \times n$ permutations by composition \cite{macwilliams1977theory}. The automorphism group of a Polar code of length $n=2^m$ is known to include at least the lower-triangular affine group LTA$(2,m)$  \cite{geiselhart2021automorphismb}. It is composed of all pairs $(\mathbf{A},\mathbf{b})$ with $\mathbf{A} \in \mathbb{F}_2^{m \times m}$ a lower-triangular matrix with unit diagonal and $\mathbf{b} \in \mathbb{F}_2^m$ a translation vector, yielding a minimum of $2^{m(m+1)/2}$ permutations. 

\subsection{Application to SBND model training}

It is easily verified that if $\mathbf{e}_\text{ML}$ is the MLD decision corresponding to input pair $(|\mathbf{y}|, \mathbf{s})$, then for any permutation $\pi \in \text{Aut}(\mathcal{C})$, $\pi(\mathbf{e}_\text{ML})$ is the MLD decision for the transformed observation $(|\pi(\mathbf{y})|, \mathbf{s'}=\pi(\mathbf{z})\mathbf{H}^t)$. Thus, by permuting the reliability vector $|\mathbf{y}|$ and the target label $\mathbf{e}_\text{ML}$, and recomputing the associated syndrome on the fly, code automorphisms provide a simple means to introduce diversity into the training examples presented to the neural decoder. The modifications to the training setup are highlighted in red in Fig. \ref{fig:sbnd}.

Note that \cite{bennatan2018deep} also exploits code automorphisms in SBND, but with a different purpose: rather than augmenting the training data, automorphisms are used at decoding time to permute each received word so that errors are concentrated in predetermined positions with high probability, thereby reducing the number of distinct error patterns the model must learn. This comes at the cost of searching over all automorphisms for each received word. In contrast, our approach uses automorphisms to expand the effective training set at negligible cost, without modifying the decoding procedure itself.

\section{Leveraging Code Automorphisms for Test-Time Augmentation at Inference}
\label{seq:test_time_aug}

The benefits of data augmentation extend beyond training. Data augmentation can also help improve prediction accuracy at inference time, a strategy known as TTA.

\subsection{Principle of test-time augmentation}

Each observation fed to the model undergoes a set of label-preserving transformations in parallel (e.g., random flips and rotations in image classification). The final prediction is obtained by aggregating the model's predictions on each transformed version. This increases the likelihood that the model will recognize learned patterns and produce correct predictions. The main drawback is higher inference cost, proportional to the number of transformations. In the following, we propose to use code automorphisms for TTA.

\begin{figure}[!t]
\centering
\includegraphics[width=88mm]{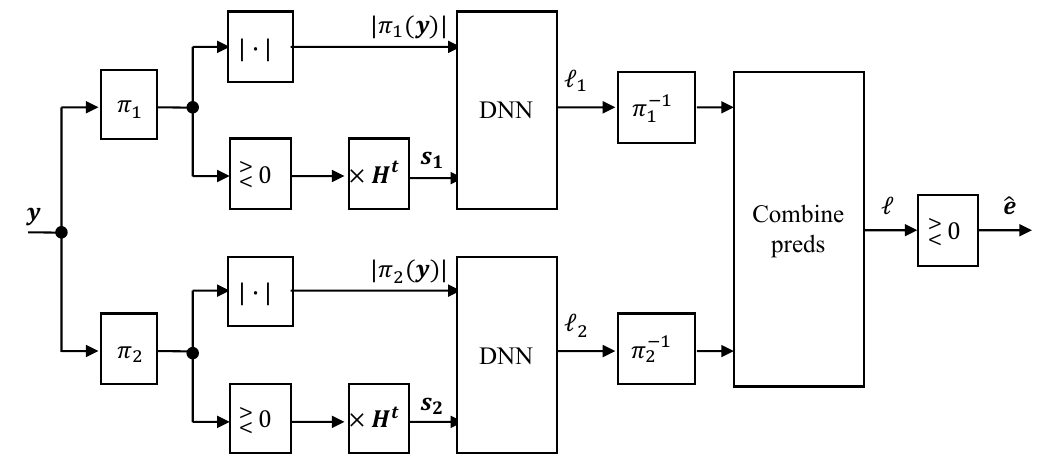}
\caption{Principle of code automorphism-based test-time augmentation.}
\label{fig:tta}
\end{figure}

\subsection{Improving SBND predictions with TTA}

Starting from an observation $(|\mathbf{y}|, \mathbf{s})$, automorphism-based TTA involves applying $P$ distinct permutations $\pi_j$ randomly selected from $\text{Aut}(\mathcal{C})$. For each transformed observation $(|\pi_j(\mathbf{y})|, \pi_j(\mathbf{z})\mathbf{H}^t)$, we compute the permuted prediction $\pi_j(\boldsymbol{\ell}_j)$ and then apply the inverse permutation $\pi_j^{-1}$ to recover $\boldsymbol{\ell}_j$. If one or more predictions yield valid codewords, we return the most likely (minimum-distance) candidate. In the absence of valid candidates, which is the most common case, we evaluated two methods to combine predictions into the final prediction $\boldsymbol{\ell}$: averaging the logits coordinate-wise, or selecting the most confident prediction (logit with maximum absolute value) for each bit. Taking the empirical average $\boldsymbol{\ell}=\frac{1}{P} \sum_{j=1}^P \boldsymbol{\ell}_j$ of the predictions proved to be the best option. The resulting inference architecture is depicted in Fig. \ref{fig:tta}. 

Interestingly, when applied to SBND, this TTA approach developed for deep learning naturally recovers the automorphism ensemble decoding technique from coding theory, which has gained renewed interest for soft decoding of short codes \cite{geiselhart2021automorphism}.

\section{Experiments}
\label{seq:experiments}

\subsection{Codes and ECCT models}

\begin{figure}[!t]
\centering
\includegraphics[width=88mm]{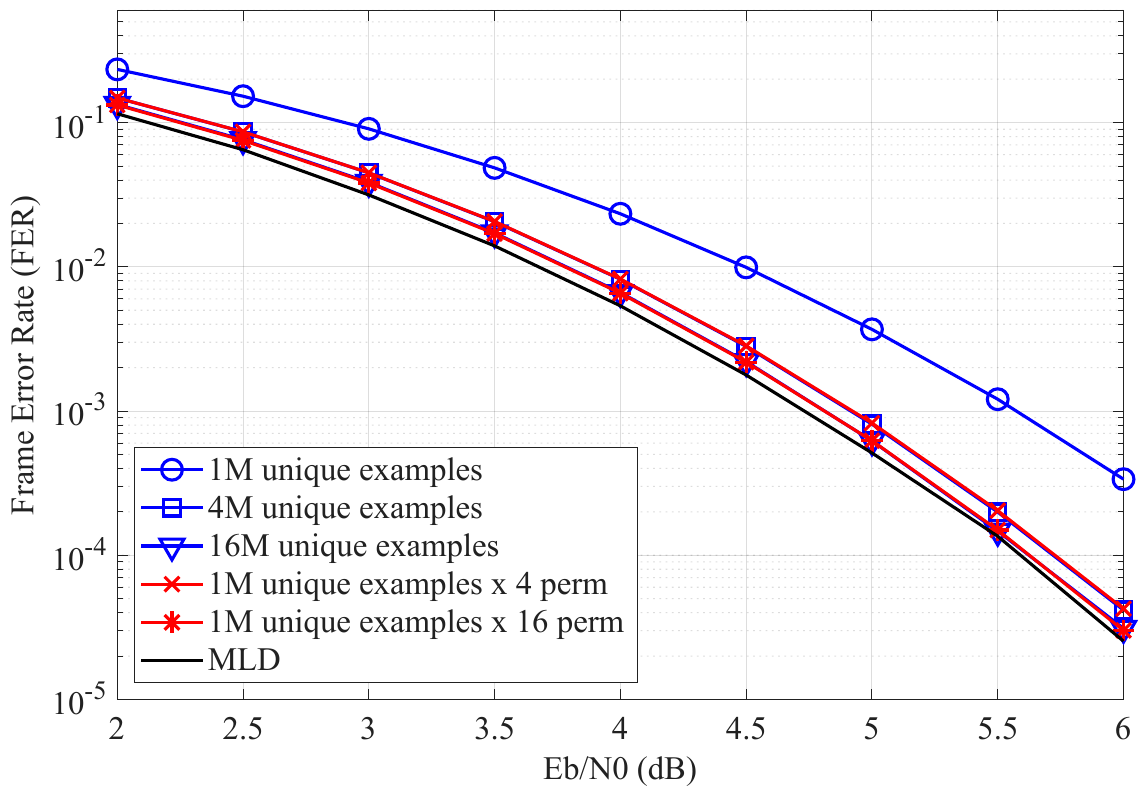}
\caption{FER for an ECCT trained to decode the $(31,21,5)$ BCH code with either unique or automorphism-augmented fixed datasets (ML labels).}
\label{fig:fer_3121_offline_aug}
\end{figure}

Results are presented for three short codes: the $(31,21,5)$ and $(63,45,7)$ BCH codes, and a $(128,64,8)$ Polar code from \cite{channelcodes}. The first is small enough to enable fast experimentation while being strong enough to provide meaningful insights. The two others have become standard baselines in the neural decoding literature, facilitating comparisons with published results. All training samples are generated at a fixed SNR value: $3$ dB for the $(31,21)$ code, $2$ dB for the $(63,45)$ code, $4$ dB for the $(128,64)$ code. This is sufficient, as models generalize well around this target value (see results below). Our ECCT implementation closely follows \cite{choukroun2022error}. All ECCT models have $h=8$ attention heads. We use $d=96$ with $N=6$ layers for the $(31,21)$ BCH code, $d=128$ with $N=6$ layers for the $(63,45)$ BCH code, and $d=128$ with $N=10$ layers for the $(128,64)$ Polar code. The corresponding models have $673$K, $1.2$M, and $2$M trainable parameters, respectively. A dropout of $0.1$ is applied to the attention weights. All experiments use 16-bit mixed-precision training with batch size $4096$, the Adam optimizer, and a warmup-stable-decay schedule with $10$ epochs for warmup and $32$ epochs for cooldown. The peak learning rate is $0.001$ for the BCH codes and $0.0001$ for the Polar code. Unless stated otherwise, models are trained for $128$ epochs. Code to reproduce all results is available at \url{https://github.com/lebidan/sbnd}.

\subsection{Measuring the training diversity of code automorphisms}
\label{sec:offline_aug}

Our first experiment aims to assess the training diversity brought by code automorphisms. Fig. \ref{fig:fer_3121_offline_aug} compares the FER performance obtained when training the ECCT to decode the $(31,21)$ BCH code on: (i) fixed datasets of 1M, 4M, and 16M unique training samples (ML labels), and (ii) augmented datasets of the same sizes created by applying 4 or 16 random permutations to each sample in the 1M base dataset. Note that augmentation occurs once during dataset construction, not during training. The model thus sees the exact same set of permuted examples repeatedly at every training epoch. Despite this, the performance reported in Fig. \ref{fig:fer_3121_offline_aug} shows no difference between the ideal case of all distinct examples and the automorphisms-augmented datasets. This remarkable result demonstrates that we can extract the same amount of training information from reduced-size datasets with the help of code automorphisms. Note also that the gap to MLD closes with increasing training set size, albeit with diminishing returns. 

\subsection{Training with data augmentation}
\label{sec:online_aug}

\begin{figure}[!t]
\centering
\includegraphics[width=89mm]{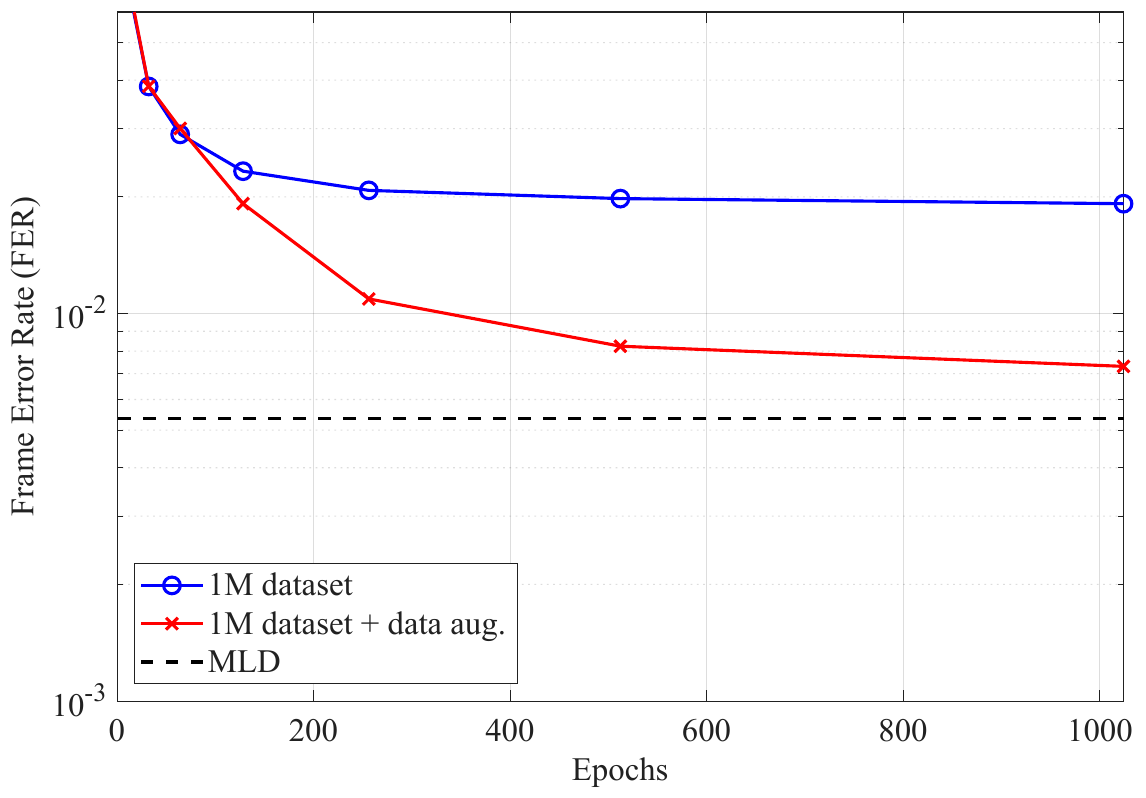}
\caption{FER at $4$ dB vs epochs, for an ECCT trained to decode the $(31,21)$ code on a 1M ML dataset w./wo automorphism-based data augmentation.}
\label{fig:evolution_online_aug}
\end{figure}

The standard deep learning practice is not to augment the dataset before training. Rather, data augmentation is performed on-the-fly during training, by applying different random transformations to each sample within each batch at every epoch. The question then arises of how to fully exploit the diversity offered by code automorphisms in this setting. Intuitively, one expects to train for more epochs to expose the model to all permutations of the training samples available, possibly multiple times for each one to improve convergence. Fig. \ref{fig:evolution_online_aug} depicts the FER evolution at $4$ dB as a function of the number of training epochs for an ECCT trained to decode the $(31,21)$ BCH code using the 1M-sample ML dataset from the previous experiment, with and without automorphism-based augmentation as described in Section~\ref{seq:train_time_aug}. Data augmentation uses the $5 \times 31 = 155$ permutations available for this code.
Without data augmentation, the model overfits the dataset after a hundred epochs and reaches a performance floor. In contrast, data augmentation leads to consistent accuracy improvement over at least a thousand epochs and achieves a final FER much closer to MLD, thereby confirming the initial intuition.

\subsection{Fixed datasets with data augmentation vs. on-demand data}
\label{sec:fixed_vs_ondemand}

\begin{figure}[!t]
\centering
\includegraphics[width=89mm]{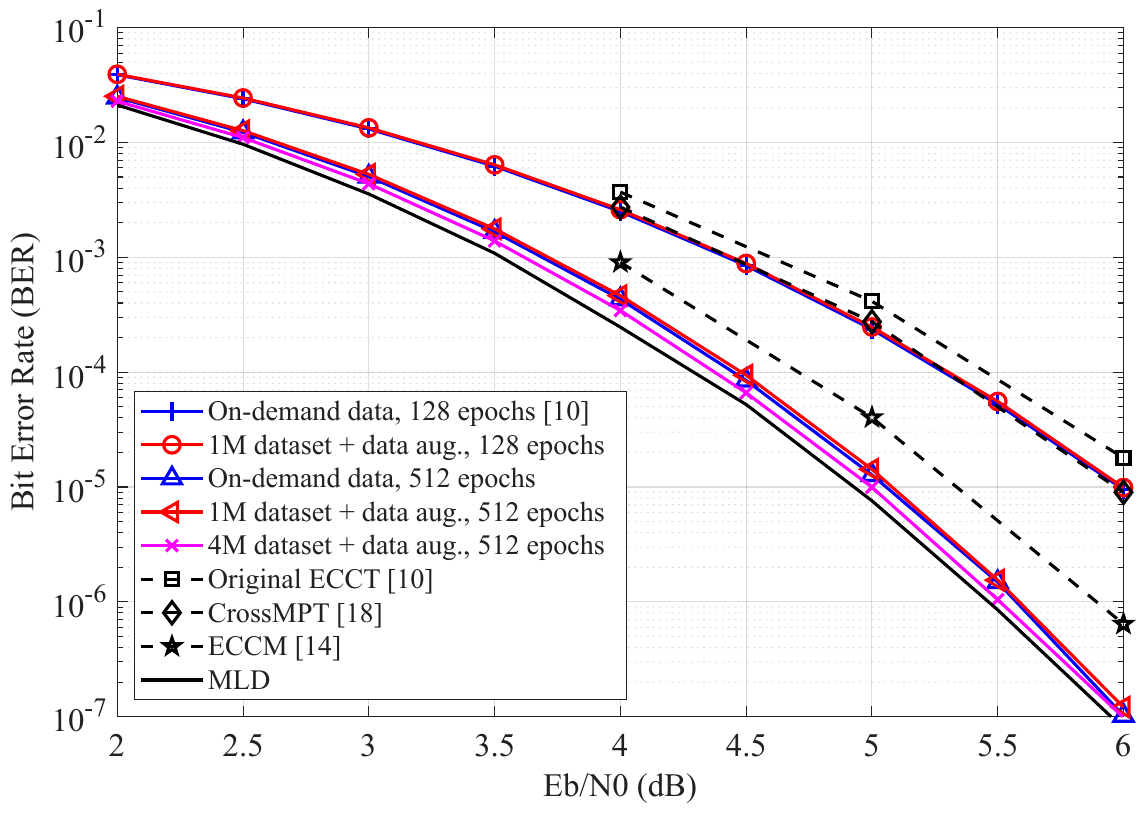}
\caption{BER performance for an ECCT trained to decode the $(63,45)$ code with oracle labels: on-demand data vs. fixed datasets with data augmentation.}
\label{fig:fixed_vs_ondemand}
\end{figure}

One may wonder whether fixed datasets with data augmentation offer any advantage over on-demand data generation. Fig.~\ref{fig:fixed_vs_ondemand} investigates both approaches when training with oracle labels for the $(63,45)$ code. We present BER performance to facilitate comparison with published results. The same ECCT model is trained with on-demand data on the one hand, and with a fixed dataset of 1M samples combined with automorphism-based data augmentation ($6 \times 63 = 378$ permutations) on the other hand. The fixed dataset is generated once before training and replayed in a different order at every epoch. For on-demand data, we use $256$ batches per epoch to match the number of training iterations with the fixed dataset (batch size $4096$). Training with on-demand oracle labels over 128 epochs reproduces the original ECCT setup from \cite{choukroun2022error}\footnote{Our results are $0.1$ dB better than those reported in \cite[Tab. 1]{choukroun2022error}, mostly due to the use of a larger batch size and a single training SNR value.}. We observe that both approaches achieve comparable performance after the same number of training steps, confirming that automorphism-based augmentation can fully compensate for the finite dataset size. More importantly, performance dramatically improves with increased training duration. At $6$ dB, after 512 epochs, the BER is reduced by $100\times$ compared to \cite{choukroun2022error}, simply by training the ECCT model four times longer. The resulting ECCT far outperforms both CrossMPT \cite[Tab. 1]{park2025crossmpt} and Mamba-Transformer \cite[Tab. 1]{cohen2025hybrid}, which appear to use the same training setup as the original ECCT. This suggests that the performance gains attributed to these newer architectures may largely reflect faster convergence rather than better decoding capability: a properly trained ECCT can match or exceed their reported results. Performance can be further improved by training for more epochs or by increasing the dataset size to 4M. Since many SBND models appear to have been trained under similar conditions as the original ECCT, these findings suggest that they are likely undertrained and that prior published results may underestimate their full potential. Note that the original GRU model of \cite{bennatan2018deep} was trained to convergence and matches the best performance in Fig.~\ref{fig:fixed_vs_ondemand}. 

\begin{figure}[!t]
\centering
\includegraphics[width=89mm]{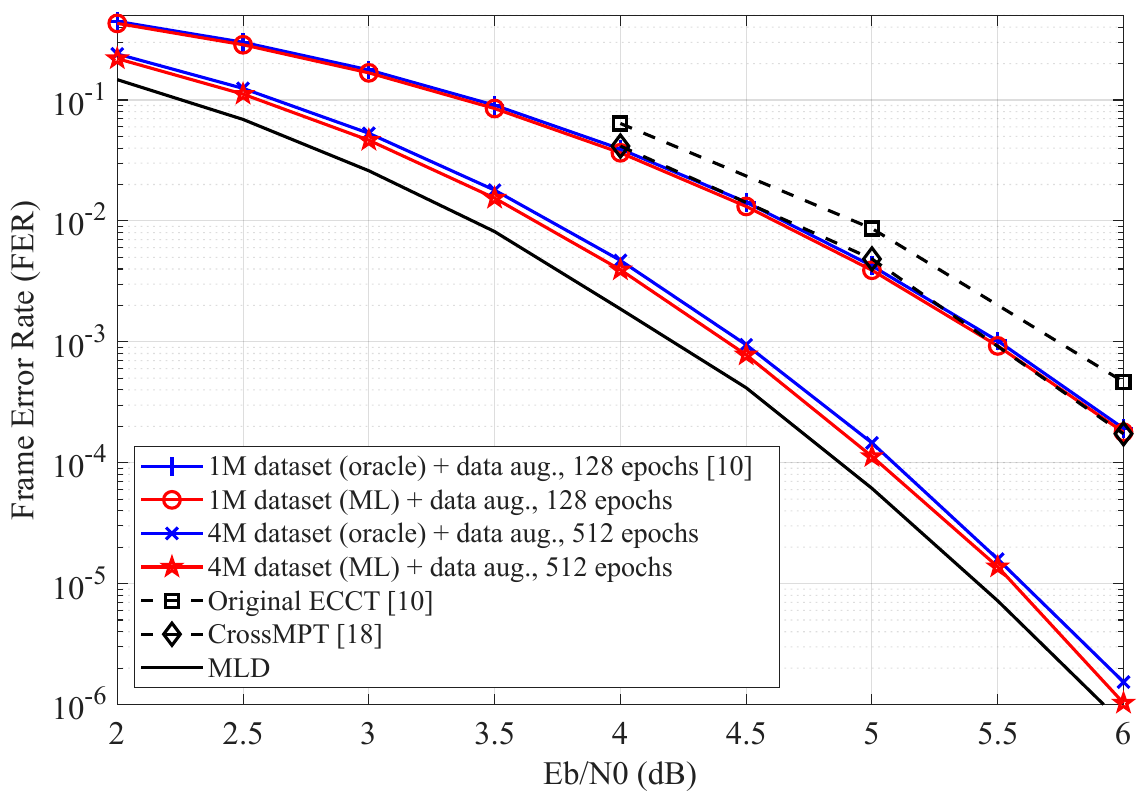}
\caption{FER performance for an ECCT trained to decode the $(63,45)$ code on a fixed dataset with augmentation: oracle labels vs. ML labels.}
\label{fig:oracle_vs_osd}
\end{figure}

\subsection{ML labels vs oracle labels}
\label{sec:label_quality}

We have seen that training on a fixed dataset augmented with code automorphisms offers no real advantage over on-demand data generation when training with oracle labels. However, replacing oracle labels with ML labels can improve accuracy for the fixed dataset approach at the same number of training epochs. This is demonstrated in Fig.~\ref{fig:oracle_vs_osd}, which compares the FER performance of an ECCT trained to decode the $(63,45)$ code with oracle and ML labels, using fixed datasets of 1M and 4M samples with data augmentation in both cases. ML labels further close the gap to MLD, and the performance advantage of ML labels over oracle labels increases with dataset size. This is because, at a fixed dataset size and training SNR value, an ML dataset differs from an oracle dataset only in a fraction $\text{FER}_\text{MLD}$ of training examples. Increasing the dataset size exposes the model to more MLD errors and brings it closer to MLD performance. Optimizing the distribution of ML labels to favor more MLD errors within the dataset can further narrow this gap without increasing dataset size \cite{ismail2025doing}. Since obtaining ML labels requires computationally expensive decoding (e.g., OSD), which is too costly to perform on the fly during training for most codes, fixed datasets are the only practical means of leveraging these high-quality labels, and automorphism-based augmentation makes it possible to do so with very compact datasets without penalizing model accuracy.

\subsection{Introducing TTA at inference time}
\label{sec:tta}

\begin{figure}[!t]
\centering
\includegraphics[width=88mm]{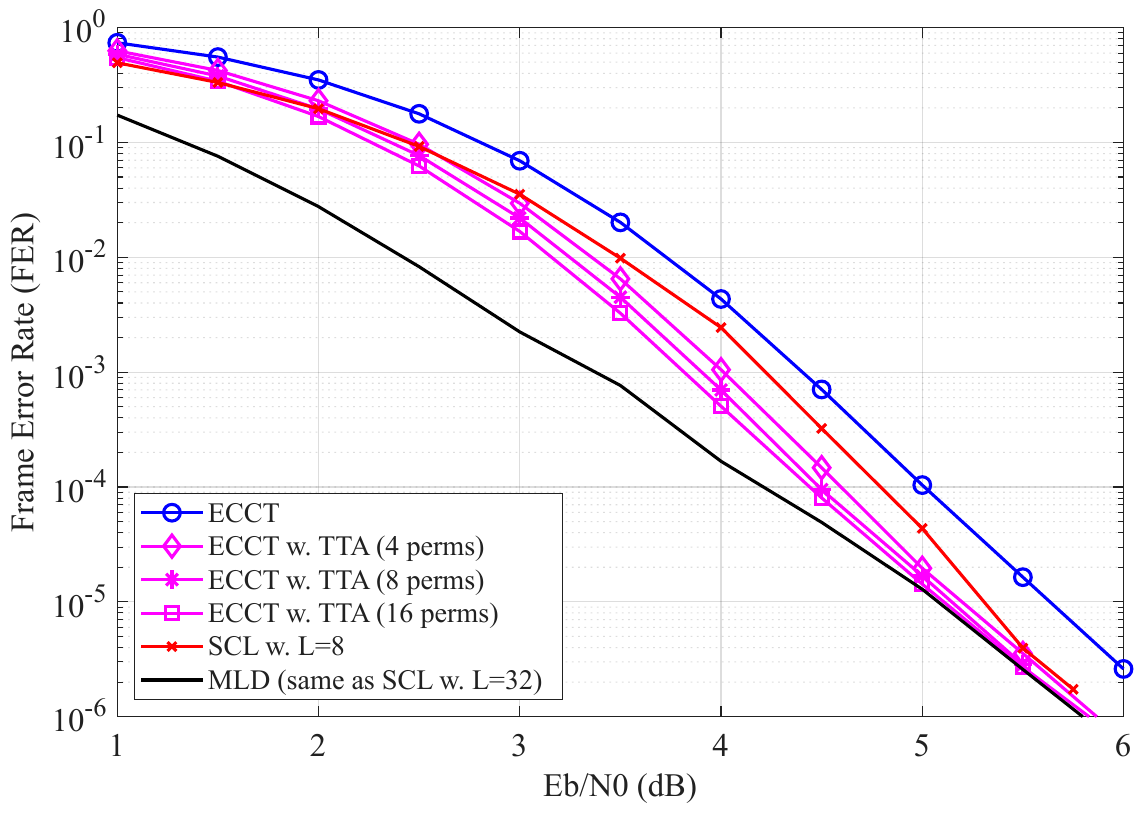}
\caption{FER performance for ECCT decoding of the $(128,64,8)$ Polar code, w./wo TTA, and comparison with Successive-Cancellation List decoding.}
\label{fig:fer_tta}
\end{figure}

Once an SBND model has been trained, with or without data augmentation, code automorphisms may still offer an additional accuracy boost through TTA. Consider the ECCT trained to decode the $(128,64,8)$ Polar code with on-demand data for 512 epochs. The FER performance of this model is shown in Fig.~\ref{fig:fer_tta} and closely matches the results reported in \cite[Fig. 4]{debonirovella2023improved}. It is only $0.2$ dB away from Successive-Cancellation List (SCL) decoding with list size $L=8$, a standard decoder baseline for Polar codes, and comes within $0.5$ dB of MLD at high SNR. Fig. \ref{fig:fer_tta} also shows the performance obtained when applying TTA to this decoder model using 4, 8, and 16 permutations, respectively. For each received word, the TTA permutations are randomly selected within a precomputed subset of LTA$(2,7)$ which is too large to be used entirely. While there is no gain in using such permutations for automorphism-SC/SCL decoding of Polar codes \cite{geiselhart2021automorphism}, the trend is very different with SBND. Introducing TTA brings noticeable performance improvement. In particular, TTA with 8 permutations matches and then progressively surpasses SCL-8 as the SNR increases, eventually reaching MLD performance at 5 dB. Using more permutations brings marginal gains. 

\section{Conclusion}
\label{seq:conclusion}

This work demonstrates that code automorphisms provide a powerful mechanism for improving SBND models through data augmentation during training and TTA at inference. The combination of fixed datasets, data augmentation, and optimized training strategies enables performance approaching MLD with limited training data. Our experimental results also indicate that some SBND models from the literature are likely undertrained and that substantial performance gains can be achieved simply through improved training methodology. Although the focus was placed on ECCT, the principles apply equally to other SBND architectures such as GRU (see, e.g. \cite{lebidan2025apport}). In addition, code automorphisms are not the only option for data augmentation; noise resampling is another direction.

We emphasize that the goal of this work is not to advocate SBND as a practical alternative to classical soft-decision decoders such as OSD, Chase, or SCL decoding. For the short high-rate codes considered here, these algorithms are likely more efficient. Rather, our results demonstrate that the performance gap between SBND models and MLD previously reported in the literature is primarily a training issue, not a fundamental limitation of these models. Whether SBND decoders can eventually match or surpass the complexity--performance trade-off of classical decoders remains an open question and an important direction for future work.

\end{document}